\documentclass[12pt]{article} \usepackage{amsmath} \usepackage{amssymb}
\usepackage{epsfig} \usepackage{verbatim} 
 \newcommand{\cC}{\mathcal{C}}
\newcommand{\cD}{\mathcal{D}} 
 \newcommand{\cM}{\mathcal{M}}

\font\openface=msbm10 at10pt
\def\Minkowski {{\hbox{\openface M}}}

\def\SetOf#1#2{\left\{ #1  \,|\, #2 \right\} }

\def\sqr#1#2{\vcenter{
  \hrule height.#2pt
  \hbox{\vrule width.#2pt height#1pt
        \kern#1pt
        \vrule width.#2pt}
  \hrule height.#2pt}}

\begin{document}

\title{Discovering the Discrete Universe}

\author{Joe Henson\footnote{Perimeter
 Institute, 31 Caroline St. N., Waterloo, N2L 2Y5, Canada}}
\maketitle

\begin{abstract}
This paper presents an brief review of some recent work on the causal set approach to quantum gravity. Causal sets are a discretisation of spacetime that allow the symmetries of GR to be preserved in the continuum approximation.  One proposed application of causal sets is to use them as the histories in a quantum sum-over-histories, \textit{i.e.} to construct a quantum theory of spacetime.

It is expected by many that quantum gravity will introduce some kind of ``fuzziness'', uncertainty and perhaps discreteness into spacetime, and generic effects of this fuzziness are currently being sought.  Applied as a model of discrete spacetime, causal sets can be used to construct simple phenomenological models which allow us to understand some of the consequences of this general expectation.
\end{abstract}
\vskip 1cm

%=====================================================

\section{Introduction: seeing atoms with the naked eye}

At present, one of the most important tasks in theoretical physics is to understand the nature of spacetime at the Planck scale.  Various indications from our current most successful theories point to this scale:  quantum effects are to be expected to invalidate the general theory of relativity here.  What should replace our current best understanding of spacetime?  This question remains controversial as no theory of quantum gravity can yet be claimed to be complete.  For example, some researchers are convinced that the kinematical structure used to replace the continuous manifolds of GR should be discrete, but others do not adhere to this requirement.  George Ellis' great contribution to our understanding of spacetime, and his interest in the issue of spacetime discreteness, make this a very appropriate topic for these proceedings.

This situation loosely parallels a well-known debate that took place somewhat over a century ago.  In that case it was not the discreteness of spacetime that was in question, but the discreteness of matter. As late as 1905 the atomic hypothesis was still doubted by figures such as Wilhelm Ostwald and Ernst Mach.  The troublesome scale in that case was around $10^{-10}$m rather than $10^{-35}$m, and even in those times, relevant experimental data was considerably less sparse that it is in the present case.  However, it is interesting to review some of the work that went into clarifying the existence of atoms of matter.  Perhaps there are some lessons to be learned for the present debate.

One of the ways to gain confidence in the atomicity of spacetime is to compare various estimates of Avogadro's number.  The convergence of disperate methods of estimation, from Loschmidt's original result in 1865 to Einstein's various determinations in 1903-1905, gave a strong reason to believe that there was something to the generic hypothesis.  Here we will concentrate on a method of Rayleigh's, published in 1899 \cite{Rayleigh:1899,Pesic:2005}\footnote{It is not entirely clear when Rayleigh developed this idea.  It was formulated in response to a letter from Maxwell sent in 1873 \cite{Pesic:2005}.  It is not generally known whether Rayleigh responded during Maxwell's lifetime.}.  This remarkable piece of physics enabled him to estimate Avogadro's number, based on a naked eye observation of Mount Everest.  His rough but well-motivated calculations and observations enabled him to understand a fundamental aspect of the nature of matter.  This stands out as a fine example of elegance in physical reasoning, and an exemplar of good physics in general.

Rayleigh used his  $\lambda^4$ scattering law to derive a relation between attenuation of light travelling thorough air and Avogadro's number.  Assuming molecules to be small, spherically symmetric spheres with negligible absorption, he derived the following:

\begin{equation}
n=\frac{32 \pi^3 (\mu - 1)^2}{3 \lambda^4 \beta}
\end{equation}

Here, $n$ is the number of molecules per cubic meter in air (related to Avogadro's number by $N_A=(2.24 \times 10^-2)n$), $\mu$ is the index of refraction of air, assumed to be close to 1 (so that $μ - 1  \ll 1$), $\lambda$ the wavelength of the incoming light (for which Rayleigh used $600$nm) and $\beta$ the scattering coefficient: light travelling through the atmosphere is extinguished by a factor of $1/e$ after travelling a length $1/\beta$.  Recalling that Everest could be seen ``fairly bright'' from Darjeeling, Rayleigh estimated $\beta \approx 160$ km.  The resulting value for Avogadro's number was of the correct order of magnitude, and roughly in agreement with the best estimates at the time.

This reasoning has a number of interesting features, which are relevant for the modern case.  The assumption of atomicity of matter was input to the model, based on various physical arguments and expectations, rather than simply output of some other theory.  It is true that the atomic hypothesis was fairly popular by 1899, but it was not forced on Rayleigh as a derived consequence of some well-developed theory (indeed, for earlier estimations, it cannot even be claimed that the atomic hypothesis had much support at the time).  Also, the calculation took place before very much was known about the dynamics of atoms.  Quantum theory was still far off.  The calculation depends only on a simple, basically kinematical, hypothesis about what molecules in air are like: they are small reflecting balls.  However, the hypothesis was not completely generic; in order to do something useful, it was not necessary to make only the most minimal assumption of atomicity.  It was only necessary that the hypothesis was simple to implement and physically well-motivated, based on the current best knowledge.

How can we apply these lessons to the problem of spacetime discreteness?  The idea of building simple models to test the hypothesis of spacetime discreteness, or some aspect of it, is intriguing.  Could analogous effects to the case of matter -- attenuation of light, random motion, \textit{etc.} -- be found in this case?  In the following section, a particular model for discrete spacetime is introduced, called the causal set \cite{Bombelli:1987aa}.  It is then explained why this particular kinematical structure is especially physically appealing.  In section \ref{s:dynamics} some ideas for using this idea as the basis for a quantum dynamics of spacetime are presented.  Finally, in section \ref{s:phenomenology} some mention is made of uses it can be put to in investigating the consequences of spacetime ``fuzziness''.  There are several other reviews on the subject of causal sets available covering different aspects of the program \cite{Wallden:2010sh,Henson:2006kf,Dowker:2006sb,Dowker:2006wr,Dowker:2005tz,Sorkin:2003bx,Sorkin:1990bh,Sorkin:1990bj}, and other reviews relevant for the motivation of causal sets \cite{Sorkin:2005qx,Sorkin:1997gi,Sorkin:1989re}.

\section{Causal Sets}
\label{s:causalsets}

Spacetime discreteness is motivated by a number of arguments in quantum gravity research, for instance, the finiteness of black hole entropy which suggests a finite number of degrees of freedom living on the surface of the black hole \cite{Sorkin:2005qx} (see \cite{Henson:2006kf} for a fuller list).  Several approaches to quantum gravity embrace some notion of discreteness.  In causal set theory, discreteness is taken as a basic hypothesis, on the strength of the physical arguments alluded to, in contrast to loop quantum gravity where discreteness is derived (not without some controversy \cite{Dittrich:2007th,Rovelli:2007ep}) as a consequence of other assumptions \cite{Sahlmann:2010zf}.
%  The asymptotic safety scenario takes an intermediate position, where, although the momentum cut-off is indeed taken to infinity, Newton's constant scales along with it, ensuring that the ratio of Planck length to the cut-off length remains finite \cite{Percacci:2007sz}.

Causal sets are a discretisation of the causal structure of continuum Lorentzian manifolds (meaning a differential manifold $\cM$ with a Lorentzian metric $g_{\mu \nu}$), \textit{i.e.}, information about which pairs of points are in each other's light-cones, and which are spacelike-related. The points of a weakly causal\footnote{A weakly causal Lorentzian manifold is one that contains no closed causal curves, otherwise called ``causal loops''.} Lorentzian manifold, together with the causal relation on them, form a partially ordered set or \textit{poset}, meaning that the set of points $C$ and the order $\prec$ on them obey the following axioms:

\paragraph{} $(i)$ Transitivity:
  $(\forall x,y,z\in C)(x\prec y\prec z\implies x\prec z).$
\paragraph{} $(ii)$ Irreflexivity:
  $(\forall x\in C)(x\not\prec x).$

\paragraph{}If $x\prec y$ then we say ``$x$ is to the past of $y$'', and if two points of the set $C$ are unrelated by $\prec$ we say they are spacelike (in short, all the normal ``causal'' nomenclature is used for the partial order).  It is interesting fact about Lorentzian geometry that ``almost all'' of the properties of a Lorentzian manifold are determined by this causal ordering.  It has been proven that, given only this order information on the points, and volume information, it is possible to find the dimension, topology, differential structure, and metric of the original manifold \cite{Hawking:1976fe,Malament:1977}.

When hypothesising spacetime discreteness, a choice must be made about which aspects of the current best description remain fundamental, and which will now only be emergent. Since the causal partial order contains so much information, it is reasonable to choose this as fundamental.  To achieve discreteness, the following axiom is introduced:

\paragraph{} $(iii)$ Local finiteness:
  $(\forall x,z\in C) \, ( {\bf card} \SetOf{y\in C}{x\prec y\prec z} < \infty ).$

\begin{figure}[ht]
\centering
\resizebox{1.4in}{0.8in}{\includegraphics{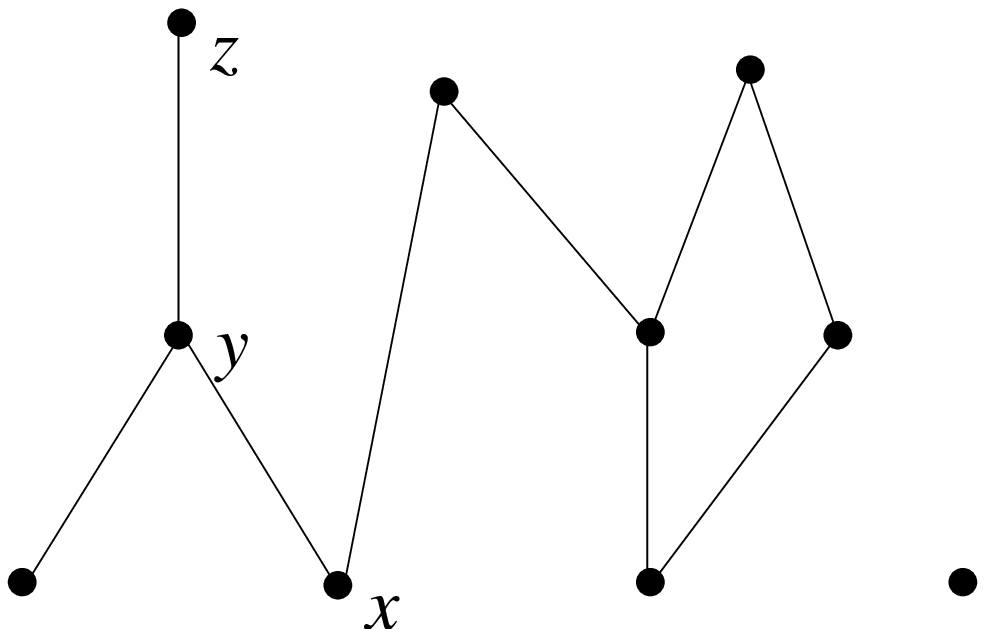}}
\caption{\small{
A causal set. The figure shows an example of a Hasse diagram.  In such a diagram, the elements of a causal set are represented by dots, and the relations not implied by transitivity (``links'') are drawn in as lines (for instance, because $x\prec y$ and $y \prec z$, there is no need to draw a line from $x$ to $z$, since that relation is implied by the other two).  The element at the bottom of the line is to the past of the one at the top of the line.
}\label{f:hasse}}
\end{figure}

\paragraph{} Where ${\bf card} \, X$ is the cardinality of the set $X$.  In other words, we have required that there only be a finite number of elements causally between any two elements in the structure (the term ``element'' replaces ``point'' in the discrete case).  A locally finite partial order is called a causal set or \textit{causet}, an example of which is illustrated in figure \ref{f:hasse}.  The \textit{causal set hypothesis} is that the appropriate description of spacetime at the Planck scale is a causal set.  Since the original results on causal structure and geometry, several researchers have independently proposed this idea \cite{Myrheim:1978ce,'tHooft:1978id,Bombelli:1987aa}.

\subsection{The Continuum Approximation}
\label{s:sprinkling}

This gives the definition of the causal set structure itself.  We now need a more precise notion of how a causal set corresponds to continuum spacetime. When can a Lorentzian manifold $(\cM,g)$ be said to be an approximation to a causet $\cC$?  Roughly, the order corresponds to the causal order of spacetime, while the volume of a region corresponds to the number of elements representing it.  But we can do a little better than this.

\begin{figure}[ht]
\centering
\resizebox{1in}{0.6in}{\includegraphics{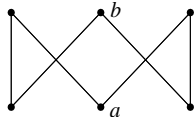}}
\caption{\small{
A Hasse diagram of the ``crown'' causet.  This causet cannot be embedded in 1+1D Minkowski space: if the above Hasse diagram is imagined as embedded into a 2D Minkowski spacetime diagram, the points at which elements $a$ and $b$ are embedded are not correctly related.  In no such embedding can the embedded elements have the causal relations of the crown causet induced on them by the causal order of 1+1D Minkowski space.  The causal set can however be embedded into 2+1D Minkowski space, where it resembles a 3 pointed crown, hence its name.
}\label{f:crown}}
\end{figure}

A causal set $\cC$ whose elements are points in a spacetime $(\cM,g)$, and whose order is the one induced on those points by the causal order of that spacetime, is said to be an \textit{embedding} of $\cC$ into $(\cM,g)$.  Not all causal sets have an embedding in every manifold\footnote{More properly, this means that not all causal sets are isomorphic to an embedded causal set in every manifold.}.  For example, the causal set in figure \ref{f:crown} cannot be embedded into 1+1D Minkowski space, but it \textit{can} be embedded into 2+1D Minkowski space.  Thus, given a causal set, we gain some information about the manifolds into which it could be embedded.  Merely requiring that a causal set embeds into an approximating manifold is not strong enough, however.  A further criterion is needed to ensure an even density of embedded elements.  This relies on the concept of sprinkling.

A sprinkling is a random selection of points from a spacetime according to a Poisson process.  The probability for sprinkling $n$ elements into a region of volume $V$ is

\begin{equation}
\label{e:poisson}
P(n)= \frac{(\rho V)^n e^{-\rho V}}{n!}.
\end{equation}

Here, $\rho$ is a fundamental density assumed to be of Planckian order.  The probability depends on nothing but the volume of the region, and so it is manifestly invariant under all volume-preserving transformations.  The sprinkling also defines an embedded causal set.  The Lorentzian manifold $(\cM,g)$ is said to approximate a causet $\cC$ \textit{if $\cC$ could have come from sprinkling $(\cM,g)$ with relatively high probability}\footnote{
The practical meaning of ``relatively high probability'' is similar to statements about the ``typicality'' of sequences of coin tosses and similar problems in probability theory. It is usually assumed that the random variable (function of the sprinkling) in question will not be wildly far from its mean in a faithfully embeddable causet.  Beyond this, standard techniques involving $\chi^2$ tests exist to test the distribution of sprinkled points for Poisson statistics.}.
  In this case $\cC$ is said to be \textit{faithfully embeddable} in $\cM$.  On average, $\rho V$ elements are sprinkled into a region of volume $V$, and fluctuations in the number are typically of order $\sqrt{\rho V}$ (a standard result from the Poisson statistics), becoming relatively insignificant for large $V$.  This gives the link between volume and number of elements.

It is important to note that the way that sprinklings and embeddings are being used here is to define a good discrete/continuum correspondence.  That is, it is necessary to use these concepts when deciding if a Lorentzian manifold approximates to a causal set.  It is not true, for example, that the fundamental structure of spacetime is supposed to be a Lorentzian manifold with a causal set embedded in it.  The causal set itself is hypothesised to completely replace all continuum spacetime structures. Of course, continuum manifolds must be approximately recovered somehow, and so it is necessary to compare discrete structures to continuum manifolds in some way (this is true, explicitly or implicitly, for any discrete replacement for spacetime).  This is the purpose that sprinkling serves, and it is the only way in which continua come into the story here.

Can such a simple structure really contain enough information to provide a good manifold approximation?  We do not want one causal set to be well-approximated by two spacetimes that are not similar on large scales.  The conjecture that this cannot happen (sometimes called the ``causal set haupvermutung'', meaning ``fundamental conjecture'') is central to the program.  It is proven in the limiting case where $\rho \rightarrow \infty$ \cite{Bombelli:1989mu}.  Also, all applications of the discrete/continuum correspondence have so far produced approximately unique values for important properties of continuum manifolds approximating to one casual set.  This gives strong evidence for the conjecture.

As an example, consider the question of dimension.  Given a fundamental causal set, how can we determine the dimension of an approximating manifold (if there is one)?  For a good estimator of dimension, it is a consequence of the haupvermutung that all manifolds that approximate to one causal set give approximately the same value.  One way to answer this is to look at the proportion of points in a causal interval (otherwise known as an Alexandrov neighbourhood, the region causally between two points) that are causally related \cite{Meyer:1988,Myrheim:1978ce,Reid:2002sj}.  In the continuum, and in flat space, it is not hard to convince oneself that one half of the pairs of points in a 2D interval are related.  The higher the dimension, the less the proportion of related points.  Inverting this relation gives the dimension as a function of the proportion of related points:

\begin{align}
D_{MM} &= f^{-1}( L ) \, , \\
f(d) &= \frac{3}{2} {\binom{\frac{3d}{2}}{d}}^{-1},
\end{align}
where $D_{MM}$ is the \textit{Myrheim-Meyer} dimension, and $L$ is the proportion of points in an interval $I$ that are related to each other.  In a causal set sprinkled into this interval, the same will be approximately true.  The proportion of pairs of sprinkled points that are related is $R/ \binom{N}{2}$ where $R$ is the total number of related pairs and $N$ is the number of points sprinkled into $I$.  Inserting this as $L$ above gives an estimate of the dimension which is accurate when $N \gg (27/16)^{D_{MM}}$.  This estimator can also be adapted to the curved-space case \cite{Reid:2002sj}.  This gives an example of how to recover effective continuum properties from the causal set structure alone.  As is clear here, the idea of sprinkling is only an intermediary used to establish the discrete/continuum correspondence, while the actual expression for the dimension only uses information intrinsic to the causal set.  Similar expressions have been found to recover distances \cite{Brightwell:1990ha,Ilie:2005qg,Rideout:2008rk}, spatial topology \cite{Major:2006hv,Major:2009cw}, and recently the scalar curvature \cite{Benincasa:2010ac}.  This gives good reason to believe that causal sets do contain enough information to pin down the continuum approximation sufficiently well.

It's important to note that some causal sets, in fact the vast majority of causal sets with a fixed large number of elements, have no continuum approximation at all \cite{Kleitman:1975}.  In this sense the existence of the continuum is not built into causal set theory at the outset.  This must be derived at a later stage (see section \ref{s:dynamics} below).

\subsection{Why this structure?}

We might now choose to employ the causal set hypothesis in different ways.  One use would be as a basis for a theory of quantum gravity, where the ``history space'' of the quantum sum-over-histories will be made up of causal sets.  A more ``Rayleighesque'' use for the causal set hypothesis is to build simple but testable models which may further inform us about what is possible and what is not when it comes to spacetime discreteness.  These uses are closely connected.  In either case the question arises: why \textit{this} kind of discreteness and not some other?  There are many ways to discretise spacetime, after all.  What singles this one out for consideration in modelling, or further, what makes it compelling as a basis for quantum gravity?

Some attractive features are already evident.  Firstly there is no barrier to sprinkling into manifolds with spatial topology change, as long as it is degeneracy of the metric at a set of isolated points that enables topology change, and not the existence of closed timelike curves (one of these conditions must exist for topology change to occur, see \textit{e.g.} \cite{Borde:1999md} and references therein).  In this discrete theory there is no problem with characterising the set of histories, as can arise in continuum path integrals.  For those who believe that topology change will be necessary in quantum gravity \cite{Sorkin:1997gi,Hawking:1978zw}, this is important.  Secondly, the structure can represent manifolds of any dimension -- no dimension is introduced at the kinematical level, as it is in Regge-type triangulations.  In fact, scale dependent dimension and topology can be introduced with the help of course-graining, as explained in \cite{Meyer:1988}, giving a natural way to deal with notions of ``spacetime foam''.  Also, it has been found necessary to incorporate some notion of causality at the fundamental level in other approaches to path integral quantum gravity, which hints towards using the causal set structure from the outset.  But the property which truly distinguishes causal sets is local Lorentz invariance \cite{Dowker:2003hb,Bombelli:2006nm}.

It may seem contradictory to claim that a discrete structure obeys a continuum symmetry.  However, this usage is carried over from condensed matter situations, and has the most physical relevance here.  We might say, for instance, that a gas, liquid or glass is rotationally invariant, as opposed to, say, a crystal.  This is relevant to the existence of fracture planes, and can affect propagation of sound and light in the medium.  What is being referred to here is the \textit{continuum approximation} to the discrete underlying structure rather than the structure itself.  In the glass, the atomic structure does not, in an of itself, serve to pick out any preferred directions in the resulting continuum approximation\footnote{One might object that the are directions, for instance the direction from each molecule to its nearest neighbour, or some function of this for every molecule.  This turns out to have no significance for any reasonable dynamics in the continuum approximation, however.  And as explained below, in the case of causal sets even this weak type of direction-picking fails.}.  Whether or not the underlying atomic configuration is rotationally invariant, or whether this question even makes sense to ask (it does not in the casual set case), is not the important point.  It is also interesting to note that the large-scale symmetry is the result of  randomness in the discrete/continuum correspondence in these cases.

Let us first treat the case of causal sets to which Minkowski space approximates, as in this case Lorentz symmetry is global and easier to examine.  It is firmly established that causal sets are Lorentz invariant in the above sense: whenever it makes sense to talk about global Lorentz symmetry, it is preserved.  This follows from three facts \cite{Bombelli:2006nm}.  Firstly, causal information is Lorentz invariant, and so taking this from a sprinkling does not bring in any frame or direction.  Secondly, the Poisson point process of sprinkling is Lorentz invariant, as the probabilities depend only on the volume and sprinkling density.  Thirdly, it has been proven that \textit{each individual instance} of the sprinkling process is Lorentz invariant.  The proof works by showing that if each sprinkling picked out a special direction, such that this picked direction transformed with the sprinkling under Lorentz transformations, this would imply that a uniform probability measure must exist on the the Lorentz group, which is impossible as the group is non-compact.  No special direction can even be associated to point in a sprinkling into Minkowski space.  Therefore, since the discrete/continuum correspondence principle given by sprinkling does not allow us to pick a direction in the approximating Minkowski space, we say that causal sets are Lorentz invariant in this sense.  In other spacetimes, the existence of local Lorentz symmetry can be claimed on similar grounds.

This is arguably the intuitive outcome: random discreteness has better symmetry properties than regular discreteness, and causal information is Lorentz invariant.  A mental picture of the sprinkled points undergoing a Lorentz transformation may tempt one to believe that there is something wrong with the idea of Lorentz invariance in this context, but this is misleading.  The analogy to glasses and crystals helps here.  To the argument that a special direction could be picked given some small region, and so Lorentz symmetry must be broken at small scales, there is an obvious answer coming from basic special relativity:  closed regions (or rulers measuring these short scales) are not Lorentz invariant.  If any physical properties can define a small region, then there is no contradiction with Lorentz symmetry if they also define a preferred direction (indeed, they must do).

Finally, phenomenological models based on causal sets manifest Lorentz invariance, helping to illustrate that the symmetry is preserved in a physically meaningful sense (see section \ref{s:phenomenology}).  This physical consideration is the root reason for calling causal sets Lorentz invariant.  Similarly, models of fields moving on lattice-like structures are well-known to violate Lorentz invariance in a continuum approximation.  The causal set, along with the discrete/continuum correspondence based on sprinkling, is the only known discretisation that can be shown to avoid this problem.  It would be possible to take more information than the causal information from the sprinkling without violating this principle, for instance the proper lengths between sprinkled points.  However, there is no reason to do this, if the causal structure and counting of points is enough to reconstruct the continuum.  In this sense, the causal set seems to be a unique discretisation.

The recovery of Lorentz Invariance in the continuum approximation is a remarkable property of causal sets.  This a significant advantage, whether we are interested in using the discrete structure to build simple heuristic models, or in constructing theories of quantum gravity \cite{Henson:2009yb}.   Considering fields moving on a discrete background, modes of any frequency can be represented equally well on a causal set; lattice-like structures, however, are well-known to violate Lorentz invariance in a continuum approximation, as they only allow a limited bandwidth of frequencies.  Assuming Planck scale discreteness leads to corrections to the diffusion relation, and other effects, which are currently under investigation.  Bounds on these effects are now threatening to overtake the range of parameters suggested by some quantum gravity inspired Lorentz violating scenarios \cite{AmelinoCamelia:2008qg,Jacobson:2005bg, Gubitosi:2010dj}.  Moreover, there is still controversy over whether introducing a Lorentz-violating cut-off in an interacting quantum theory can be made anywhere near consistent with observation, without (at best) severe fine tuning \cite{Collins:2004bp,Perez:2003un,Visser:2009fg}.  If we do not find these scenarios entirely compelling, and even more so if they become ruled out, it is as well to have alternative models.  Besides this, maintaining the well-established principle of local Lorentz symmetry has the advantage of severely limiting the possible types of discreteness.

\subsection{Non-locality and causal sets}

\begin{figure}[ht]
\centering
\resizebox{4.5in}{2.3in}{\includegraphics{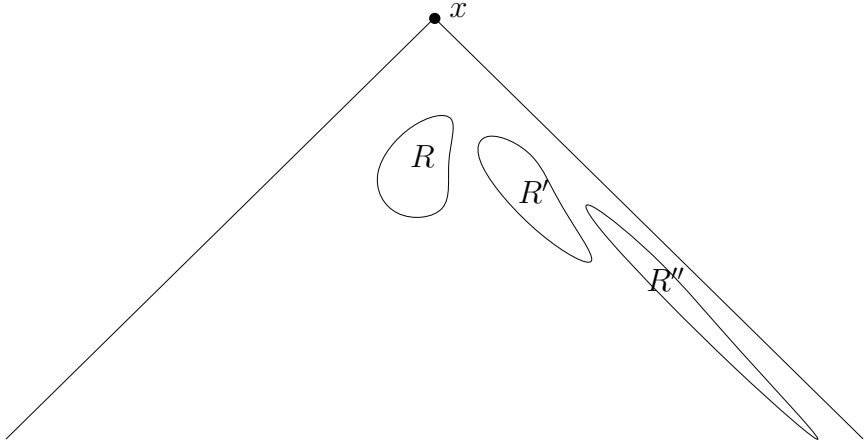}}
\caption{\small{
In this diagram, $x$ is a point in a sprinkling of 2D Minkowski space, with its past light-cone.  $R$ is a region of this space, $R'$ is a region found by boosting $R$, and $R''$ is found by boosting $R'$ by the same factor.  There are an infinite series of these disjoint regions, each of which has the same probability for containing a sprinkled element that qualifies as a nearest neighbour of $x$ (by whatever Lorentz invariant criterion) .  Because of this, the number of neerest neighbours must be infinite, however they are defined.
}\label{f:lorentz}}
\end{figure}

These good symmetry properties of causal sets are linked to non-locality.  In lattices, each vertex is linked to a finite number of nearest neighbours, and this ``locality'' makes it easy to discretise operators like the Laplacian in this context.  This is not so easy for causal sets.  Consider a point $x$ in a sprinkling of Minkowski space, as illustrated in figure \ref{f:lorentz}.  How can we define ``nearest neighbours'' in a way that is intrinsic to the causal set?  It could for instance be a sprinkled point $y$ ``linked'' to the past of $x$, meaning that $y \prec x$ and there are no elements $z$ such that $y \prec z \prec x$.  These linked elements must be close to $x$ in a faithful embedding, otherwise the probability for there to be an ``in-between'' element like $z$ becomes large, as we can see from equation (\ref{e:poisson}). You could imagine other definitions. Whatever they are, we have to pick the nearest neighbours in a Lorentz invariant way (in particular we cannot mark any other point in the spacetime).  Let us try to find them to the past.  Whatever the criterion for being a nearest neighbour, there will be some probability $p$ for there to be one or more nearest neighbours in some finite volume region $R$ close to $x$ (if the probability was 0 for any such region it wouldn't be a useful definition).  But, if we boost $R$ by a large enough factor, we can find another region $R'$, disjoint from $R$.  Due to the Lorentz invariance of sprinkling, the probability of finding some nearest neighbours in $R'$ is also $p$.  Continuing this to $R''$, $R'''$, \textit{etc.}, we find an infinite series of regions, each with some finite probability to contain nearest neighbours.  Thus, the number of nearest neighbours is infinite.

This is does not come about because the information in the casual set is too sparse; it would occur even if we took more information from the sprinkling to determine our discrete structure.  The argument above is quite general for any form of discreteness based on sprinklings which respects Lorentz invariance.  This shows how non-locality is the flip-side of Lorentz invariant discreteness.  This isn't surprising, considering that there is an infinite volume within any finite proper time of a point in Minkowski space, due to the non-compactness of the Lorentz group.  A finite number of nearest neighbours would have to lie near some particular direction away from the point $x$.  We see in particular that the number of links from $x$ is infinite, and linked elements occur in a narrow band that extends all the way down the light cone, in a sprinkling of Minkowski space.

Despite this non-locality, recent work has manged to recover good approximations to local operators on causal sets by various means \cite{Sorkin:2007qi, Benincasa:2010ac}, showing that this non-locality is not a terminal problem.  This works suggests that there may be a new length scale intermediate between the discreteness scale and macroscopic scales, at which locality breaks down.  These ideas might have interesting consequences for phenomenology, as discussed below.

\section{Towards quantum gravity}
\label{s:dynamics}

As yet there is no quantum dynamics for causal sets that might serve as a basis for a theory of quantum gravity.  However, recent progress suggests that may soon be possible to write down a such a theory and test it using simulations.

\subsection{Growth models}

There are two main approaches to causal set quantum dynamics.  One, perhaps the most discussed to date, is to choose a general dynamical framework for causal set dynamics, and then constrain it by the use of some physical principles.  As a stepping stone to the quantum case, a stochastic dynamics has been formulated called Classical Sequential Growth dynamics \cite{Rideout:1999ub,Varadarajan:2005gg}.  Here, the dynamical framework is one of ``growth'', where causal set elements are added to the future and spacelike to existing elements, with some probability attached to each alternative way of doing so.  These probabilities are then constrained to obey principles of general covariance and local causality \textit {\`a la} Bell.  The result is a dynamics with one free parameter per causet element, which, however, has several generic features for large classes of parameters.

This model has given rise to several lines of work.  One task is to characterise the physical questions that the theory can answer, in keeping with the symmetry of general covariance, which in this context means labelling invariance.  This is the causal set analog of a ``problem of time'' as seen in canonical theories.  The problem here can be solved, in the sense that the class of meaningful questions can be found and given a simple and physically appealing interpretation \cite{Brightwell:2002vw,Dowker:2005gj}.

The model also gives some interesting results for cosmology, and the emergence of the continuum.  Some generic models have a ``bouncing cosmology'' with many big-bang to big-crunch cycles.  The large spatial extent of the universe in these models is not a result of fine-tuning, but simply a consequence of the large age of the universe, giving a mechanism for fixing parameters that may be useful in more realistic theories \cite{Sorkin:1998hi,Martin:2000js,Ash:2002un}.  The CSG models have also been of use in developing tests of dynamically generated causal sets to look for manifoldlike behaviour \cite{Rideout:2000fh}, and computational techniques for causal sets.  It has been shown that manifoldlike causal sets are not typical outcomes for the CSG models at largest scales \cite{Brightwell:2010}.  At these scales the causal set has a basically 1-dimensional character.  Recently, however, it was found that at some intermediate range of scales, some properties of CSG-generated causets were found to match the properties expected for de Sitter space \cite{Ahmed:2009qm}.  It seems probable at this point, though not definite, that the model will remain as a useful example on the way to a quantum sequential growth model, rather than giving GR as an approximation in its own right.  Still, the model continues to provide a testing ground for new ideas and to give some hints about what the full theory might look like.

\subsection{Generalising the path integral}

Another approach, which has until recently been less well studied, is closer to the approach taken in other quantum gravity programs such as that based on Causal Dynamical Triangulations \cite{Ambjorn:2009ts}.  In this approach, a quantum dynamics is sought that is similar in form to the path integral for standard quantum theories.  Roughly, one seeks to generalise the path integral for transition probabilities to the gravitational case, replacing particle trajectories or field configurations with the chosen space of discrete structures, and using an appropriate action and measure factor.

Taking a more foundational view, this picture might be too limited for causal sets, or in fact any quantum gravity theory.  Asking only about transition probabilities implies asking about states on spatial slices, which is not such a natural concept in a generally covariant theory; most interesting observables, for example various measures of the effective dimension for causal sets or Regge-type discretisations, are not of this type.  Indeed, for a causal set very little information is given by the analog of the ``configuration on a spatial slice''.  The corresponding concept for a causal set is a subset of causally unrelated elements, that is maximal, \textit{i.e.} to which no further spacelike elements can be added.  The only information in such a ``slice'' is the number of elements it contains!  This shows that the causal set is a type of discretisation in which a spacetime formulation is particularly natural, rather than a ``space and time'' canonical one.  To see how spacetime relates to the causal set one must look beyond a single slice; a causal set derived from sprinkling is ``non-local in time'' in this sense.

A generalised framework is therefore useful.  Quantum theory can of course make predictions about events that refer to properties of the history at more than one time, and all these probabilities are captured in an object called the quantum measure \cite{Sorkin:1994dt}, or equivalently the decoherence functional \cite{Hartle:1992as} (for some recent progress on these ideas, see \cite{Dowker:2010ng}).  This object can also be given a path integral formulation for standard theories, and it is this that we seek to generalise to the gravitational case.

A toy model suggests what might be possible here.  In 2D, there is an easily identifiable subclass of causal sets that contains all causal sets to which intervals of 2D Lorentzian spacetimes can approximate (and some non-manifoldlike causal sets besides) called the ``2D orders''.  We can make this class our history space, and supply an appropriate action which gives the continuum 2D action for causal sets that are well approximated by spacetimes (which turns out to be trivial in this case).  With this done, the suprising result is that sprinklings of flat space dominate the path integral \cite{Brightwell:2007aq}.  As we have seen, sprinklings were introduced to casual set theory for physical reasons, to make sure that the discrete/continuum approximation had good symmetry properties.  It is remarkable and encouraging, therefore, that they also appear naturally in this setting.

However, this intriguing 2D result may not be easy to generalise to a sum over all causal sets.  It was based on a number of theorems that are only known for the 2D orders. Also, the action is trivial in this case.  A natural choice for the action is a function of the causal set that approximates to the integral of the scalar curvature for any approximating 4D spacetime.  It had until recently been difficult to construct such local quantities.  Now, however, due to new work by Dionigi Benincasa and Fay Dowker, an expression for the action has been discovered \cite{Benincasa:2010ac}.  The action is derived from discretisations of the d'Alembertian operator on the causal set.  It uses the concept of $n$-element inclusive order intervals, which are subsets of the causal set comprising \textit{all} the elements causally between some pair of element, which have $n$-elements \textit{including} that pair. A 2-element inclusive order interval is just a link.  From the correspondence between volume and number, order intervals with small numbers of elements represent small regions in any corresponding spacetime, although they could be in any frame, \textit{i.e.} you could find two that were highly boosted with respect to one another.  The action for 4D is given by
\begin{equation}
\label{e:action}
\frac{1}{\hbar} S(\cC) = N - N_1 + 9N_2 - 16N_3 + 8N_4,
\end{equation}
where $N$ is the number of elements in the causal set $\cC$ and $N_i$ is the number of $(i + 1)$-element inclusive order intervals in $\cC$.  The lack of parameters in the action is a consequence of setting the bare Planck scale to be 1, and the bare cosmological constant $\Lambda$ (which would simply add a term proportional to $\Lambda N$) to 0.  The particular sequence of coefficients follows from more detailed considerations; in this respect the situation is similar to the  discretisation of the second derivative for a function $f$ on a lattice with separation $\Delta x$.  A standard expression in this case is $f(x- \Delta x) - 2f(x) + f(x + \Delta x)$, where the sequence of coefficients $\{1,-2,1\}$ is clearly necessary to find the right continuum limit.

A remaining problem involves actually calculating Lorentzian path integrals.  Writing down such an expression may now be possible, but this is not much use if no interesting results can be calculated.  The great strength of the CDT approach is that observables such as the scaling effective dimension, and the spatial extent of the universe as a function of time, can be calculated using Monte Carlo simulations.  This is achieved by analytically continuing a parameter that multiplies the time.  The path integral can then be Wick rotated: the Lorentzian geometries are transformed to Euclidean ones, and in the process Feynman's $e^{iS/\hbar}$ turns into a more Boltzmanian $e^{-S/\hbar}$, which can then be dealt with by the well-developed methods of statistical physics.

In the case of casual sets, the idea of Wick rotation has no obvious analogue.  However, analytical continuation may still be employed to calculate path integrals.  In the the case of a non-relativistic particle, instead of Wick rotating, the mass parameter can be analytically continued to solve the path integral, and similar methods are relevant for quantum gravity \cite{Sorkin:2009ka}.  The same kinds of ideas allow a statistical path integral for causal sets to be derived from the Feynman path integral, by analytically continuing an appropriate parameter which can be included in the action.  Finding some Euclideanised analog of a causal set is therefore unnecessary.  This, along with the new expression for the action, opens the door for Monte Carlo simulations in causal set theory analogous to those that have been performed in CDT theory.  It would be surprising if summing over all causal sets with the action above produced a semi-classical state dominated by well-behaved 4D spacetimes; after all, this sum includes approximate spacetimes with all dimensions and topologies, and a large number of other causal sets besides these.  However, this kind of work could be the beginning of a process of refinement much like that which led to CDT theory from previous models, progressing by limiting the history space or altering the action.  And, as explained below, we can see that at least a ``worst case scenario'' can be avoided, giving hope that a well-behaved continuum approximation may be possible.

\subsection{Problems with entropy, solutions from non-locality}

\begin{figure}[ht]
\centering
\resizebox{2.3in}{0.8in}{\includegraphics{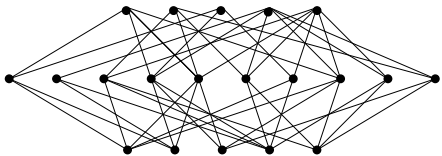}}
\caption{\small{
A typical Kleitman-Rothschild causal set with 20 elements.  As the number of elements becomes large, this type of causal set numerically dominates all others.
}\label{f:kr}}
\end{figure}

As alluded to above, most causal sets do not resemble manifolds.  In fact, for large number of elements $N$, the vast majority of causal sets have a particular and very non-manifoldlike structure \cite{Kleitman:1975}.  A ``Kleitman-Rothschild'' causal set (otherwise called a KR order in the mathematical literature) is composed of three layers, with approximately $N/4$, $N/2$ and $N/4$ elements respectively, as shown in figure \ref{f:kr}.  The elements of the top layer are to the future of some of those in the middle layer, the elements in the bottom layer are to the past of some of the elements in the middle layer, and all of those in the bottom layer are to the past of all of those in the top layer.  Taking a causal set uniformly at random from the space of all causal sets (which you might call the ``purely entropic'' or ``unweighted sum-over-causets'') typically produces this kind of structure.

Considering such a structure as a discrete spacetime, we have a universe with a very short duration!  In a certain sense these structures are also infinite dimensional.  Consider a set of discrete spacetimes with varying $N$ that represent the same geometry with a different overall scale (this also applies to triangulations \textit{etc.}). For a $d$-dimensional discrete spacetime, we expect the total time duration to scale like $N^{1/d}$ and the size of the maximal spatial surface to scale like $N^{(d-1)/d}$ \footnote{This interpretation only makes sense when the same large-scale geometry, up to scale, is seen as $N$ varies.  For example, if the structure represented a long chain of successive 4D universes separated by cosmological big-crunch-big-bang ``bounces'', this would give a scaling dimension of 1.}.  By both these measures KR orders have $d=\infty$.  Interestingly, this is similar to the situation in both dynamical triangulations \cite{Ambjorn:1995aw} and causal dynamical triangulations  \cite{Ambjorn:2009ts}, where a purely entropic measure produces a highly connected, infinite-dimensional ``crumpled state''.  It's encouraging that this can be seen analytically for causal sets (at least in the case of the ``unrestricted'' sum over \textit{all} causal sets).

The number of KR orders grows like $2^{N^2/4}$ and quickly comes to vastly outnumber all others.  The proportion of causal sets that have a height of greater than 3 scales like $C^{-N}$ for some constant $C$ \cite{Brightwell:1996}.  Although it is hard to put a figure on the total number of manifoldlike causal sets for each value of $N$, it is expected that the scaling of the proportion of ``bad'' configurations to ``good'' ones would be even worse.

These ``bad'' histories cannot be the ones that dominate any physically reasonable path integral or ``sum over histories''.  The configurations that dominate a statistical sum-over-histories arise from an interplay between the entropy of different types of history, and their actions.  When the proportion of ``bad'' configurations grows faster than exponentially with the number of elements or vertices in a discrete structure, this may set alarm bells ringing, for the following reason.  It's tempting to think that, if the action is a sum of local terms, it should scale with the total size of the system $N$ for a typical history, and thus the weight $e^{-S}$ would scale exponentially with $N$.  If that is true for bad configurations and good ones alike, we can see that the ratio between a bad and good weight can at best scale exponentially, while the proportion of bad configurations grows super-exponentially.  Such an action cannot therefore prevent the bad configurations from dominating.  This problem occurs, for example, if one allows all spacetime topologies in a Euclidean sum over 2D triangulations.

However, in the case of causal sets, we can see that this worry is unfounded.  The action will not scale in this way for manifoldlike causal sets\footnote{It is interesting to note that this would not be the case for any graph-based discreteness, in which manifoldlike graphs have a low average valency that does not scale with $N$, and where the action is based on graph-local quantities like valency.  It is however still true that the number of edges in a typical random graph, which would be non-manifoldlike, scales faster than $N$.  This leaves the option of cutting some bad configurations out of the history space.} or for Kleitman-Rothschild causal sets.  This is because the number of links, and other such structures, in these kinds of causal sets do not scale like $N$.

Let us consider the KR orders.  The first term in the action in equation (\ref{e:action}) is simply $N$, but the second term in the action is the number of links. For any pair made up of one top layer element and one middle layer element, there may or may not be a link.  The number of possible links between top and middle is therefore $(N/4)(N/2)$.  The same is true between the bottom and middle layers giving a total of $N^2/4$ possible links.  For a typical K-R order about half of these will be present, and so the second term in the action scales like $N^2$, not $N$ as in the argument above\footnote{This argument treats the elements as if they are distinguishable, or labelled, which we do not want if we are respecting the discrete analog of general covariance; we really want to talk of ``unlabelled causal sets'' which are isomorphism classes of labelled causal sets.  However, it is not unnatural to weight each unlabelled causal set $C$ by a ``measure factor'' of $1/|Aut(C)|$ where $|Aut(C)|$ is the number of automorphisms of $C$, and in this case the problem dissapears.  Also, it can be shown that in almost all KR orders each element is indeed distinguished from all others by the pattern of relations, \textit{i.e.} they have no non-trivial automorphisms.}.  It  can also be seen that the other terms in equation (\ref{e:action}) are small for a typical KR order.  Taking a bottom layer element and a top layer element, the number of elements causally between them is the number of middle layer elements that are linked to both.  This will be around $N/8$ for a typical KR order, and so the number of $3$, $4$ and $5$-element inclusive order intervals will scale much more slowly than $N$.  Thus the second term dominates, and (because the sign of this term in the action will remain positive after our analytic continuation procedure\footnote{More details of this process will be given a future publication with David Rideout, Rafael Sorkin and Sumati Surya.}) will suppress typical KR orders by a term of the form $e^{-N^2/4}$.  Thus, the non-locality of causal sets is a positive feature in this case, as the action can do more to suppress bad configurations than a strictly local action could in standard cases.

This does not by itself imply that manifoldlike causal sets will dominate, but it does dispatch the argument that the faster-than-exponential growth in the non-mainfoldlike configurations is an insurmountable problem.  It is not yet known exactly how the action scales for a typical sprinkling into flat space; the average over sprinklings is known to be 0 (neglecting boundary terms), which is encouraging, but the size of the fluctuations also has to be taken into account.  This could be settled by simulations of sprinklings. At least, we do have the 2D result to show that, in some statistical sums, sprinklings can arise naturally.  A resolution to the question of whether they can be selected by a full sum over causal sets awaits the implementation of Monte-Carlo simulations, which are now being developed.

\section{Consequences of spacetime Discreteness}
\label{s:phenomenology}

Having discussed some more ambitious uses for causal sets, we can now return to our original theme: developing simple models based on physically appealing hypotheses, and then testing them against observation and experiment.  We might first consider generalising standard models, simply by replacing Minkowski space with a corresponding causal set.  This helps to test if causal set discreteness is consistent with present observations, and also to look for new effects. Using the causal set hypothesis in this way fulfils some goals brought up in the introductory discussion.  As in Rayleigh's case, this approach does not depend on the form of the dynamics for causal sets; it is essentially a non-quantum modification, which, like the atomic hypothesis for matter, might reveal some interesting consequences for discreteness before the quantum dynamics is known.  Also like the 100-year-old case, these kinds of models posit discreteness at the outset, as a hypothesis to be tested, rather than waiting for it to be derived from some fully satisfying theory.  It is an input from physical considerations, rather than an output from mathematical considerations.  The physical considerations in this case are the arguments pointing to discreteness, coupled with the requirement of Lorentz symmetry.  Models based on causal sets therefore contrast with most work on the phenomenology of quantum gravity, which concentrates on the possibility of Lorentz violation.  This alone would give good motivation to study such models.

Finally, we assume something more than the most generic possible hypothesis, that is, that quantum gravity will lead to some type of fuzziness, uncertainty or discreteness in spacetime, which will show up at or around the Planck scale.  On the one hand, the more generic the hypothesis, the more significant the result might be for various approaches to quantum gravity. On the other hand some specificity is usually necessary to actually construct a model, as Rayleigh's case illustrates.  In any case, such specific models can lead to generic conclusions in one way: by providing counter-examples to generic claims.  And of course any positive observations would be just as significant as for more generic models, if not more so.

One remarkable case from causal set theory is the successful prediction -- the only such prediction to date from quantum gravity research -- of the order of magnitude of the cosmological constant using a heuristic argument based on the general features of a causal set quantum dynamics \cite{Sorkin:1990bh,Ahmed:2002mj,Sorkin:2005fu}.  This was a genuine prediction, in the sense that it was made at a time when observations were consistent with $\Lambda=0$.  This reasoning gives rise to a scenario is which the cosmological ``constant'' is subject to fluctuations, something that could give rise to further predictions.

Another idea that has been investigated is a spacetime analog of Brownian motion.  Arguments have been put forward that suggest that particles moving on a causal set background may not follow geodesics, but instead might deviate from them slightly as a result of spacetime discreteness \cite{Dowker:2003hb}.  The resulting model allows a type of Lorentz invariant diffusion of the energy-momentum of particles\footnote{This point has given rise to some confusion in the literature.  The fact that the model does not respect momentum or energy conservation does not imply a breaking of Poincar\'e symmetry.  This is because the usual assumptions of the Noether theorem are not respected in this case -- the underlying dynamics is not based on a local Lagrangian, and the resulting process is diffusive.  Adding some kind of energy bath to the model could conceivably bring it back into the Lagrangian framework, but in this case, the lack of energy conservation in the subsystem outside the bath is no suprise either.}.  The magnitude of the corresponding ``diffusion constant'' has not been derived from fundamental considerations; indeed the model is only heuristic and contains assumptions that might not be true in a more detailed dynamics.  However, the model is useful in that it can be tested.  Bounds can be set on the diffusion constant by various laboratory experiments and astronomical observations \cite{Dowker:2003hb, Kaloper:2006pj}.  Similar models introduce Lorentz invariant diffusion into the motion of massless particles, and consider the case of polarisation, comparing results to observations of the CMBR \cite{Philpott:2008vd,Contaldi:2010fh}.

Another line of work concentrates on putting field theory onto a background causal set.  The idea is to have a field theory, described intrinsically in terms of the causal set, that approximates to the standard case when the causal set corresponds to a continuum spacetime.  As mentioned above, the discretisation of the d'Alembertian operator has been achieved, allowing a classical scalar field dynamics to be defined \cite{Sorkin:2007qi, Benincasa:2010ac}.  Also, significant steps towards a quantum field theory on a causal set background have been taken.  Most important amongst these are the definition of Green's functions for massive scalar fields \cite{Johnston:2008za}, including a definition of the Feynman propagator on the causal set \cite{Johnston:2009fr}.  Work to fully define such a field theory is in progress.  This would not only give an opportunity to search for new effects, but also provide a new, Lorentz invariant way of imposing a cut-off on quantum field theory.

There is another way to treat fields moving on a causal set background, that is simple enough to describe briefly here\footnote{This method seems not to give rise to momentum diffusion effects.}. This will shed some light on the issue of the coherence of light from distant sources.

\subsection{A discrete model of wave propagation}

It has been claimed on fairly general grounds that any ``fuzziness'' in spacetime due to quantum gravity effects will lead to a loss of coherence of light from distant sources.  If we replace Minkowski space with a causal set derived from sprinkling that spacetime, we do introduce uncertainties in spacetime properties, albeit only on the kinematical level; lengths between points, for example, can only be reconstructed to some finite accuracy.  Would a model of propagation of light on this spacetime therefore fall victim to this problem?  Using certain assumptions, some authors have derived effects that have already been ruled out by observation \cite{Lieu:2003ee}; others argue against these assumptions \cite{Amelino-Camelia:2004yd, Coule:2003td, Ng:2003ag,Christiansen:2005yg}, but also find in their models that some types of spacetime uncertainty are ruled out.  Looking closely at the assumptions made in either case, we can see that the uncertainty is of a Lorentz violating form: uncertainties are considered separately for the wavelength and frequencies of light.  Therefore we can already see that these arguments would not apply to a causal set model.  But the question of whether causal set discreteness is consistent with the coherence of light from distant sources is not yet answered by these considerations.  This can be tested in a specific model which is briefly described here (more details can be found in \cite{Dowker:2010aa}).

For these purposes, the essential features of the situation can be modeled with a massless scalar field propagating from a source to a detector. The source represents some distant astronomical source.  In this case, the relevant dynamics of the scalar field are summarised in the retarded Green's functions, which in (3+1)D flat space is a delta function on the forward light-cone:

\begin{align}
  \label{Green} G(y,x) &= \begin{cases} \frac{1}{2 \pi} \delta(|y -
  x|^2) & \text{if $y$ is in the causal future of $x$} \\ 0 &
  \text{otherwise} \\ \end{cases} \\ &= \frac{1}{4 \pi r} \delta (y^0
  - x^0 - r),
\end{align}
where $r$ is the spatial distance from $x$ to $y$.  In terms of $G$,
the field produced by our source is given by
\begin{gather}
  \phi(y) = \int_P \, G(y,x(s)) \, q \, ds \ ,
\end{gather}
where $P$ is the worldline of the source, $q$ is its charge, and $s$
is proper time along $P$.  We can model the detector signal $F$ as the integrated field in some small and distant detector region $\cD$, giving
\begin{gather}
\label{e:Fdoubleint}
  F = q \int_P \, ds \int_{\cD} \, d^4y \, G(y,x(s)).
\end{gather}

From the form of the Green's function, we can see that the signal is basically just a measure of the pairs of null-related points, one of which is in the source, and the other of which is to the future of it in the detector (it is the volume of such pairs the relevant space, $\Minkowski^4 \times \Minkowski^4$).  With certain approximations, the result for a point charge is

\begin{equation}
\label{e:continuumoutput}
        F \approx (1+v) \; \frac{q}{4 \pi R} \;
        V \ ,
\end{equation}
where $V$ is the spacetime volume of the detector, $R$ is the spatial separation of source and detector, and $v$ is the relative velocity of the source to the detector.  Combining a moving negative and stationary positive source, to cancel the constant term in (\ref{e:continuumoutput}), gives a result that has the same form as the electromagnetic case.  We can make $v=a sin(\omega t)$ for some angular frequency $\omega$ and amplitude $a$ (for simplicity we take the duration of detection to be smaller that the period of oscillation here, making the detector more than realistically accurate for most applications).  The phase of the source is reflected in the detector output, and so observations in keeping with this model demonstrate the coherence of light from distant sources.

Now we want to discretise this model.  We replace the 4D Minkowski space by a causal set, to which Minkowski space is only an approximation.  According to our rule introduced above, this is a causal set generated from a typical sprinkling into Minkowski space.  As per usual for discretisations of fields, the field $\phi$ is becomes a real number on each causal set element.  The main difficulty is how to discretise the Green's functions.  We need a function that, in the limit, gives non-zero values only on the forward light-cone.  To this end, consider the causet function
\begin{equation}
\label{e:ldef}
L(x,y)=
\begin{cases} \kappa & \text{whenever $x \prec y$ and $\{x,y\}$ is a link,}
\\
0 &\text{otherwise,}
\end{cases}
\end{equation}
where $x,y \in \cC$ are causet elements and $\kappa$ is a normalising
constant of order 1, that we set to make the correspondence to the continuum model correct.  We have already seen that links to any element lie very close to the light-cone and extend all the way along it.  In the limit of infinitely dense sprinkling, scaling $\kappa$ appropriately, this function does indeed become a $\delta$-function on the forward light-cone of $x$.  The simple causal character of the Green's functions has thus made it very natural and easy to discretise them on the causal set.

With these definitions in place, we can proceed to repeat the continuum calculation.  In that case, the signal output was proportional to a volume measure of all pairs of points, one in the source and the other in the detector, that were future-null related.  In the causal set case, this simply becomes the number of future directed links between the source and the detector.  With suitable definitions of the source and detector region, the calculation can be performed.  For a typical sprinkling into Minkowski space, the result turns out to the the same as the continuum one, with a small discrepancy overwhelmingly due to the random fluctuations inherent to sprinkling.  One coherent source, which has been mentioned in the context of loss of coherence due to quantum gravity effects, is Active Galactic Nucleus PKS 1413+135, which is at a distance of order one Gigaparsec.  With this distance, the discrepancy between the continuum and causal set models turns out to be about one part in $10^{12}$, even for unrealisticly stringent paremeters, and some modelling assumptions that tend to increase fluctuations.  The coherence of light from distant sources is thus preserved in this model.

The conclusion here is that causal set discreteness does not cause problems for the coherence of light from distant sources.  This implies that the generic idea of introducing Planck-scale ``uncertainties'' or ``fluctuations'' of some otherwise unspecified type does not inevitably lead to loss of coherence.  The result also justifies the initial expectation based on considerations of Lorentz invariance.  This is good in one way, but dissappointing in another;  the ideal situation is to make a prediction of a small discrepancy from standard theory that can then be searched for, rather than one many orders of magnitude below what can be detected.  That is another reason for pursuing the models of energy and momentum diffusion mentioned above, and for more detailed consideration of the ``fluctuating cosmological constant'' scenario.

\section{Conclusion: back to the rough ground}

This review started with a look at how a simple physical hypothesis about discreteness can, with some thought, be developed into a model that gives considerable insight.  In the 19th century, the idea of atomic matter developed from a natural speculation to a more compelling proposition, based on hints from the best theories of the day.  Arguments were put forward that combined understanding from these theories (like electromagnetism) with simple hypotheses (such as that atoms were hard or or perfectly reflecting spheres) to develop testable models.  Early on, some researchers criticised the introduction of such simplistic hypotheses\footnote{Kelvin scorns the ``assumption of infinitely strong and infinitely rigid pieces of matter'' as ``monstrous'' in a paper on the vortex theory of matter \cite{Kelvin:1867}.}, which seemed to them to be unjustified.  Instead, they urged that the atomicity of matter should follow from some more elegant generalisation of the current theories. The most notable case is Kelvin's ``vortex atom'' theory \cite{Kelvin:1867}, in which atoms were vortices in a hypothesised perfect liquid -- a subject that was well-studied at the time, and undergoing many interesting mathematical advances.  However, the more basic hypotheses, leading as they did to clear physical consequences (as in Rayleigh's case), turned out to have a much greater influence on the formation of the new theory.

It must be said that this is only one way of reading the history, and it does not prove that the same will be true in the case of spacetime discreteness.  But, at least, it motivates the application of the same approach in this case.  Because of the difficulty of obtaining relevant data, efforts to formulate a discrete model with clear phenomenological relevance are likely to be more challenging than they were previously.  Despite this, there are several approaches to probing quantum gravity and spacetime discreteness with the aid of heuristic arguments, and beyond this, to setting up a theory of quantum gravity based on the hypothesis of discrete spacetime.

In this spirit, spacetime discreteness has been investigated above.  The fact that geometry can be almost totally described in terms of the causal relation motivates the idea that it could be the causal structure that is most fundamental, and survives discretisation (rather than, for example, distance relations).  Coupling the discreteness hypothesis with another physically motivated requirement, that the symmetries of GR be preserved in the continuum approximation, gives a related way to arrive at the causal set hypothesis.

There are several open avenues of research in causal set theory.  There are always more ``kinematical'' questions to be answered, relating to the discrete/continuum correspondence: can we, for example, approximately deduce the metric of a spatial hypersurface in an approximating manifold, by considering only information intrinsic to the corresponding causal set?  And how large are the fluctuations in the new estimator for the scalar curvature, in flat space for example?  After the recent success in defining the Feynman propagator for scalar fields living on a causal set, the problem of fully defining a scalar quantum field theory is of great interest and is currently under study.  Defining a dynamics for vector and spinor fields would also be useful for phenomenological models.  As for the dynamics of causal sets themselves, the calculation of some ``discretised path integrals'' is also now within reach, as reviewed above, promting the development of Monte Carlo simulations.  This means that, for the first time, a quantum dynamics for causal sets could be studied, and, if simulations are practical, some physically interesting results could be derived from it.  Also on the question of dynamics, work on the idea of sequential growth is continuing, with some studies of a possible quantum generalisation of the CSG models.

For phenomenology, further study of the heuristic derivation of the value of the cosmological constant is also an interesting area for more research.  Along with this, there is the idea of looking at the consequences of the causal set action, which, due to causal set non-locality, would lead to some small corrections to the standard expression in the continuum approximation \cite{Benincasa:2010ac}.  There are still many avenues to be explored concerning fields moving on a causal set background: for instance, to see if any new effects follow from the new discretisation of the d'Alembertian operator.  Similar work could also proceed in the quantum case, once the relevant theory is established.  With these techniques there is some hope of making contact with data from astronomical observations, particularly with cosmology, the area that currently seems most promising as a provider of relevant data for quantum gravity research.  These lines of research hopefully bring us closer to the goal of ``seeing'' the discreteness of spacetime in our best current observations, as Rayleigh managed to see the discreteness of matter with the naked eye.

\paragraph{}The author is grateful to the organisers and participants of the ``Foundations of Space and Time'' conference held in Cape Town in 2009, celebrating George Ellis' birthday, which was a lively and interesting meeting.  Thanks are also due to participants in the Causal Sets '09 conference at DIAS, Dublin, where some of the new ideas of section \ref{s:dynamics} were developed and discussed, and to Peter Pesic for some interesting information about Rayleigh's observation.  Research at Perimeter Institute for Theoretical Physics is supported in part by the Government of Canada through NSERC and by the Province of Ontario through MRI.

\bibliographystyle{h-physrev3} %% [[RDS  trying this out]]
\bibliography{refs.bib}

\begin{thebibliography}{10}

\bibitem{Rayleigh:1899}
Rayleigh,
\newblock On the transmission of light through an atmosphere containing small
  particles in suspension,
\newblock in {\em Scientific Papers by Lord Rayleigh} Vol.~4, pp. 247--405, New
  York: Dover, 1899/1964.

\bibitem{Pesic:2005}
P.~Pesic,
\newblock Eur. J. Phys. {\bf 26}, 183 (2005).

\bibitem{Bombelli:1987aa}
L.~Bombelli, J.-H. Lee, D.~Meyer, and R.~Sorkin,
\newblock Phys. Rev. Lett {\bf 59}, 521 (1987).
%%CITATION = PRLTA,59,521;%%

\bibitem{Wallden:2010sh}
P.~Wallden,
\newblock (2010), 1001.4041.
%%CITATION = 1001.4041;%%

\bibitem{Henson:2006kf}
J.~Henson,
\newblock {The causal set approach to quantum gravity},
\newblock in {\em {Approaches to Quantum Gravity: Towards a New Understanding
  of Space and Time}}, edited by D.~Oriti, Cambridge University Press, 2006,
  gr-qc/0601121.
%%CITATION = GR-QC/0601121;%%

\bibitem{Dowker:2006sb}
F.~Dowker,
\newblock Contemp. Phys. {\bf 47}, 1 (2006).
%%CITATION = CTPHA,47,1;%%

\bibitem{Dowker:2006wr}
F.~Dowker,
\newblock AIP Conf. Proc. {\bf 861}, 79 (2006).
%%CITATION = APCPC,861,79;%%

\bibitem{Dowker:2005tz}
F.~Dowker,
\newblock (2005), gr-qc/0508109.
%%CITATION = GR-QC/0508109;%%

\bibitem{Sorkin:2003bx}
R.~D. Sorkin,
\newblock (2003), gr-qc/0309009.
%%CITATION = GR-QC/0309009;%%

\bibitem{Sorkin:1990bh}
R.~D. Sorkin,
\newblock First steps with causal sets,
\newblock in {\em Proceedings of the ninth Italian Conference on General
  Relativity and Gravitational Physics, Capri, Italy, September 1990}, edited
  by R.~Cianci {\em et~al.}, pp. 68--90, World Scientific, Singapore, 1991.

\bibitem{Sorkin:1990bj}
R.~D. Sorkin,
\newblock Space-time and causal sets,
\newblock in {\em Relativity and Gravitation: Classical and Quantum,
  Proceedings of the SILARG VII Conference, Cocoyoc, Mexico, December 1990},
  edited by J.~C. D'Olivo {\em et~al.}, pp. 150--173, World Scientific,
  Singapore, 1991.

\bibitem{Sorkin:2005qx}
R.~D. Sorkin,
\newblock Stud. Hist. Philos. Mod. Phys. {\bf 36}, 291 (2005), hep-th/0504037.
%%CITATION = HEP-TH/0504037;%%

\bibitem{Sorkin:1997gi}
R.~D. Sorkin,
\newblock Int. J. Theor. Phys. {\bf 36}, 2759 (1997), gr-qc/9706002.
%%CITATION = GR-QC 9706002;%%

\bibitem{Sorkin:1989re}
R.~D. Sorkin,
\newblock (1989), gr-qc/9511063.
%%CITATION = GR-QC 9511063;%%

\bibitem{Dittrich:2007th}
B.~Dittrich and T.~Thiemann,
\newblock J. Math. Phys. {\bf 50}, 012503 (2009), 0708.1721.
%%CITATION = 0708.1721;%%

\bibitem{Rovelli:2007ep}
C.~Rovelli,
\newblock (2007), 0708.2481.
%%CITATION = 0708.2481;%%

\bibitem{Sahlmann:2010zf}
H.~Sahlmann,
\newblock (2010), 1001.4188.
%%CITATION = 1001.4188;%%

\bibitem{Hawking:1976fe}
S.~W. Hawking, A.~R. King, and P.~J. McCarthy,
\newblock J. Math. Phys. {\bf 17}, 174 (1976).
%%CITATION = JMAPA,17,174;%%

\bibitem{Malament:1977}
D.~B. Malament,
\newblock J. Math. Phys. {\bf 18}, 1399 (1977).

\bibitem{Myrheim:1978ce}
J.~Myrheim,
\newblock CERN-TH-2538.

\bibitem{'tHooft:1978id}
G.~'t~Hooft,
\newblock Talk given at 8th Conf. on General Relativity and Gravitation,
  Waterloo, Canada, Aug 7-12, 1977.

\bibitem{Bombelli:1989mu}
L.~Bombelli and D.~A. Meyer,
\newblock Phys. Lett. {\bf A141}, 226 (1989).
%%CITATION = PHLTA,A141,226;%%

\bibitem{Meyer:1988}
D.~Meyer,
\newblock PhD thesis, MIT, 1988.

\bibitem{Reid:2002sj}
D.~D. Reid,
\newblock Phys. Rev. {\bf D67}, 024034 (2003), gr-qc/0207103.
%%CITATION = GR-QC 0207103;%%

\bibitem{Brightwell:1990ha}
G.~Brightwell and R.~Gregory,
\newblock Phys. Rev. Lett. {\bf 66}, 260 (1991).
%%CITATION = PRLTA,66,260;%%

\bibitem{Ilie:2005qg}
R.~Ilie, G.~B. Thompson, and D.~D. Reid,
\newblock (2005), gr-qc/0512073.
%%CITATION = GR-QC 0512073;%%

\bibitem{Rideout:2008rk}
D.~Rideout and P.~Wallden,
\newblock Class. Quant. Grav. {\bf 26}, 155013 (2009), 0810.1768.
%%CITATION = 0810.1768;%%

\bibitem{Major:2006hv}
S.~Major, D.~Rideout, and S.~Surya,
\newblock J. Math. Phys. {\bf 48}, 032501 (2007), gr-qc/0604124.
%%CITATION = GR-QC/0604124;%%

\bibitem{Major:2009cw}
S.~Major, D.~Rideout, and S.~Surya,
\newblock Class. Quant. Grav. {\bf 26}, 175008 (2009), 0902.0434.
%%CITATION = 0902.0434;%%

\bibitem{Benincasa:2010ac}
D.~M.~T. Benincasa and F.~Dowker,
\newblock (2010), 1001.2725.
%%CITATION = 1001.2725;%%

\bibitem{Kleitman:1975}
D.~Kleitman and B.~Rothschild,
\newblock Trans. Amer. Math. Society {\bf 205}, 205 (1975).

\bibitem{Borde:1999md}
A.~Borde, H.~F. Dowker, R.~S. Garcia, R.~D. Sorkin, and S.~Surya,
\newblock Class. Quant. Grav. {\bf 16}, 3457 (1999), gr-qc/9901063.
%%CITATION = GR-QC 9901063;%%

\bibitem{Hawking:1978zw}
S.~W. Hawking,
\newblock Nucl. Phys. {\bf B144}, 349 (1978).
%%CITATION = NUPHA,B144,349;%%

\bibitem{Dowker:2003hb}
F.~Dowker, J.~Henson, and R.~D. Sorkin,
\newblock Mod. Phys. Lett. {\bf A19}, 1829 (2004), gr-qc/0311055.
%%CITATION = GR-QC 0311055;%%

\bibitem{Bombelli:2006nm}
L.~Bombelli, J.~Henson, and R.~D. Sorkin,
\newblock (2006), gr-qc/0605006.
%%CITATION = GR-QC 0605006;%%

\bibitem{Henson:2009yb}
J.~Henson,
\newblock J. Phys. Conf. Ser. {\bf 174}, 012020 (2009), 0901.4009.
%%CITATION = 0901.4009;%%

\bibitem{AmelinoCamelia:2008qg}
G.~Amelino-Camelia,
\newblock (2008), 0806.0339.
%%CITATION = 0806.0339;%%

\bibitem{Jacobson:2005bg}
T.~Jacobson, S.~Liberati, and D.~Mattingly,
\newblock (2005), astro-ph/0505267.
%%CITATION = ASTRO-PH 0505267;%%

\bibitem{Gubitosi:2010dj}
G.~Gubitosi, G.~Genovese, G.~Amelino-Camelia, and A.~Melchiorri,
\newblock (2010), 1003.0878.
%%CITATION = 1003.0878;%%

\bibitem{Collins:2004bp}
J.~Collins, A.~Perez, D.~Sudarsky, L.~Urrutia, and H.~Vucetich,
\newblock Phys. Rev. Lett. {\bf 93}, 191301 (2004), gr-qc/0403053.
%%CITATION = GR-QC 0403053;%%

\bibitem{Perez:2003un}
A.~Perez and D.~Sudarsky,
\newblock Phys. Rev. Lett. {\bf 91}, 179101 (2003), gr-qc/0306113.
%%CITATION = GR-QC 0306113;%%

\bibitem{Visser:2009fg}
M.~Visser,
\newblock Phys. Rev. {\bf D80}, 025011 (2009), 0902.0590.
%%CITATION = 0902.0590;%%

\bibitem{Sorkin:2007qi}
R.~D. Sorkin,
\newblock (2007), gr-qc/0703099.
%%CITATION = GR-QC/0703099;%%

\bibitem{Rideout:1999ub}
D.~P. Rideout and R.~D. Sorkin,
\newblock Phys. Rev. {\bf D61}, 024002 (2000), gr-qc/9904062.
%%CITATION = GR-QC 9904062;%%

\bibitem{Varadarajan:2005gg}
M.~Varadarajan and D.~Rideout,
\newblock (2005), gr-qc/0504066.
%%CITATION = GR-QC 0504066;%%

\bibitem{Brightwell:2002vw}
G.~Brightwell, H.~F. Dowker, R.~S. Garcia, J.~Henson, and R.~D. Sorkin,
\newblock Phys. Rev. {\bf D67}, 084031 (2003), gr-qc/0210061.
%%CITATION = GR-QC 0210061;%%

\bibitem{Dowker:2005gj}
F.~Dowker and S.~Surya,
\newblock (2005), gr-qc/0504069.
%%CITATION = GR-QC 0504069;%%

\bibitem{Sorkin:1998hi}
R.~D. Sorkin,
\newblock Int. J. Theor. Phys. {\bf 39}, 1731 (2000), gr-qc/0003043.
%%CITATION = GR-QC 0003043;%%

\bibitem{Martin:2000js}
X.~Martin, D.~O'Connor, D.~P. Rideout, and R.~D. Sorkin,
\newblock Phys. Rev. {\bf D63}, 084026 (2001), gr-qc/0009063.
%%CITATION = GR-QC 0009063;%%

\bibitem{Ash:2002un}
A.~Ash and P.~McDonald,
\newblock J. Math. Phys. {\bf 44}, 1666 (2003), gr-qc/0209020.
%%CITATION = GR-QC 0209020;%%

\bibitem{Rideout:2000fh}
D.~P. Rideout and R.~D. Sorkin,
\newblock Phys. Rev. {\bf D63}, 104011 (2001), gr-qc/0003117.
%%CITATION = GR-QC 0003117;%%

\bibitem{Brightwell:2010}
G.~Brightwell and N.~Georgiou,
\newblock Random Struct. Algorithms {\bf 36}, 218 (2010).

\bibitem{Ahmed:2009qm}
M.~Ahmed and D.~Rideout,
\newblock (2009), 0909.4771.
%%CITATION = 0909.4771;%%

\bibitem{Ambjorn:2009ts}
J.~Ambjorn, J.~Jurkiewicz, and R.~Loll,
\newblock (2009), 0906.3947.
%%CITATION = 0906.3947;%%

\bibitem{Sorkin:1994dt}
D.~Sorkin, Rafael,
\newblock Mod. Phys. Lett. {\bf A9}, 3119 (1994), gr-qc/9401003.
%%CITATION = GR-QC 9401003;%%

\bibitem{Hartle:1992as}
J.~B. Hartle,
\newblock Space-time quantum mechanics and the quantum mechanics of space-time,
\newblock in {\em Proceedings of the Les Houches Summer School on Gravitation
  and Quantizations, Les Houches, France, 6 Jul - 1 Aug 1992}, edited by
  J.~Zinn-Justin and B.~Julia, North-Holland, 1995, gr-qc/9304006.
%%CITATION = GR-QC 9304006;%%

\bibitem{Dowker:2010ng}
F.~Dowker, S.~Johnston, and R.~D. Sorkin,
\newblock (2010), 1002.0589.
%%CITATION = 1002.0589;%%

\bibitem{Brightwell:2007aq}
G.~Brightwell, J.~Henson, and S.~Surya,
\newblock Class. Quant. Grav. {\bf 25}, 105025 (2008), 0706.0375.
%%CITATION = 0706.0375;%%

\bibitem{Sorkin:2009ka}
R.~D. Sorkin,
\newblock (2009), 0911.1479.
%%CITATION = 0911.1479;%%

\bibitem{Ambjorn:1995aw}
J.~Ambjorn, J.~Jurkiewicz, and Y.~Watabiki,
\newblock J. Math. Phys. {\bf 36}, 6299 (1995), hep-th/9503108.
%%CITATION = HEP-TH/9503108;%%

\bibitem{Brightwell:1996}
G.~Brightwell, H.~J. Pr{\"o}mel, and A.~Steger,
\newblock J. Comb. Theory, Ser. A {\bf 73}, 193 (1996).

\bibitem{Ahmed:2002mj}
M.~Ahmed, S.~Dodelson, P.~B. Greene, and R.~Sorkin,
\newblock Phys. Rev. {\bf D69}, 103523 (2004), astro-ph/0209274.
%%CITATION = ASTRO-PH 0209274;%%

\bibitem{Sorkin:2005fu}
R.~D. Sorkin,
\newblock Braz. J. Phys. {\bf 35}, 280 (2005), gr-qc/0503057.
%%CITATION = GR-QC 0503057;%%

\bibitem{Kaloper:2006pj}
N.~Kaloper and D.~Mattingly,
\newblock Phys. Rev. {\bf D74}, 106001 (2006), astro-ph/0607485.
%%CITATION = ASTRO-PH/0607485;%%

\bibitem{Philpott:2008vd}
L.~Philpott, F.~Dowker, and R.~D. Sorkin,
\newblock Phys. Rev. {\bf D79}, 124047 (2009), 0810.5591.
%%CITATION = 0810.5591;%%

\bibitem{Contaldi:2010fh}
C.~Contaldi, F.~Dowker, and L.~Philpott,
\newblock (2010), 1001.4545.
%%CITATION = 1001.4545;%%

\bibitem{Johnston:2008za}
S.~Johnston,
\newblock Class. Quant. Grav. {\bf 25}, 202001 (2008), 0806.3083.
%%CITATION = 0806.3083;%%

\bibitem{Johnston:2009fr}
S.~Johnston,
\newblock Phys. Rev. Lett. {\bf 103}, 180401 (2009), 0909.0944.
%%CITATION = 0909.0944;%%

\bibitem{Lieu:2003ee}
R.~Lieu and L.~W. Hillman,
\newblock Astrophys. J. {\bf 585}, L77 (2003), astro-ph/0301184.
%%CITATION = ASTRO-PH 0301184;%%

\bibitem{Amelino-Camelia:2004yd}
G.~Amelino-Camelia,
\newblock (2004), gr-qc/0402009.
%%CITATION = GR-QC 0402009;%%

\bibitem{Coule:2003td}
D.~H. Coule,
\newblock Class. Quant. Grav. {\bf 20}, 3107 (2003), astro-ph/0302333.
%%CITATION = ASTRO-PH 0302333;%%

\bibitem{Ng:2003ag}
Y.~J. Ng, H.~van Dam, and W.~A. Christiansen,
\newblock Astrophys. J. {\bf 591}, L87 (2003), astro-ph/0302372.
%%CITATION = ASTRO-PH 0302372;%%

\bibitem{Christiansen:2005yg}
W.~A. Christiansen, Y.~J. Ng, and H.~van Dam,
\newblock Phys. Rev. Lett. {\bf 96}, 051301 (2006), gr-qc/0508121.
%%CITATION = GR-QC/0508121;%%

\bibitem{Dowker:2010aa}
F.~Dowker, J.~Henson, and R.~D. Sorkin,
\newblock Discreteness and the transmission of light from distant sources,
\newblock in preparation.

\bibitem{Kelvin:1867}
Kelvin,
\newblock Proceedings of the Royal Society of Edinburgh {\bf VI}, 94 (1867),
\newblock \\ http://zapatopi.net/kelvin/papers/on\_vortex\_atoms.html.

\end{thebibliography}
%% \bibliography{working.bbl}

%%% \bibliographystyle{h-physrev3}
%%% \bibliography{../Bibliography/refs}

\end{document}